\documentclass[conference]{IEEEtran}
\IEEEoverridecommandlockouts
\ifCLASSINFOpdf
   \usepackage[pdftex]{graphicx}
\else
\fi
\usepackage{amsmath}
\usepackage{amssymb}  

\usepackage{url}

\usepackage{etoolbox}
\makeatletter
\patchcmd{\@makecaption}
  {\scshape}
  {}
  {}
  {}
\makeatother

\usepackage{color}
\definecolor{vermillion}{rgb}{0.86, 0.18, 0.01}

\newcommand{\sis}{\sqrt{\text{iSWAP}}}
\begin{document}
%
\title{Large scale multi-node simulations of $\mathbb{Z}_2$ gauge theory quantum circuits using Google Cloud Platform}

\author{
\and
\IEEEauthorblockN{
Erik Gustafson\IEEEauthorrefmark{1} \\ and
Burt Holzman\\ and
James Kowalkowski\\ and
Henry Lamm\\ and
Andy C. Y. Li\\ and
Gabriel Perdue \\}
\IEEEcompsocitemizethanks{\IEEEcompsocthanksitem\IEEEauthorrefmark{1} Corresponding author: egustafs@fnal.gov}
\IEEEauthorblockA{Fermi National Accelerator Laboratory\\
Batavia, IL 60510}
\and
\IEEEauthorblockN{
Sergio Boixo \\ and
Sergei V. Isakov\\ and
Orion Martin\\ and
Ross Thomson\\ and
Catherine Vollgraff Heidweiller \\}
\IEEEauthorblockA{Google\\
Mountain View, CA 94043}
\and
\IEEEauthorblockN{
Jackson Beall\\ and
Martin Ganahl\\ and
Guifre Vidal \\}
\IEEEauthorblockA{Sandbox@Alphabet\\
Mountain View, CA 94043}
\and
\IEEEauthorblockN{Evan Peters \IEEEauthorrefmark{2}  }
\IEEEcompsocitemizethanks{\IEEEcompsocthanksitem{\IEEEauthorrefmark{2} Corresponding author: e6peters@uwaterloo.ca } }
\IEEEauthorblockA{Institute for Quantum Computing\\
University of Waterloo\\
Waterloo, Ontario N2L 3G1, Canada}
}


%


\maketitle

\begin{abstract}
Simulating quantum field theories on a quantum computer is one of the most exciting fundamental physics applications of quantum information science. Dynamical time evolution of quantum fields is a challenge that is beyond the capabilities of classical computing, but it can teach us important lessons about the fundamental fabric of space and time.
Whether we may answer scientific questions of interest using near-term quantum computing hardware is an open question that requires a detailed simulation study of quantum noise.
Here we present a large scale simulation study powered by a multi-node implementation of \emph{qsim} using the Google Cloud Platform. 
We additionally employ newly-developed GPU capabilities in \emph{qsim} and show how Tensor Processing Units --- Application-specific Integrated Circuits (ASICs) specialized for Machine Learning --- may be used to dramatically speed up the simulation of large quantum circuits.
We demonstrate the use of high performance cloud computing for simulating $\mathbb{Z}_2$ quantum field theories on system sizes up to 36 qubits.
We find this lattice size is not able to simulate our problem and observable combination with sufficient accuracy, implying more challenging observables of interest for this theory are likely beyond the reach of classical computation using exact circuit simulation.
\end{abstract}


%
\IEEEpeerreviewmaketitle

\section{Introduction}

Quantum field theory (QFT) on the lattice ultimately requires us to extrapolate to the infinite volume and continuum limits. This extrapolation requires a large number of highly connected qubits, and the ability to check systems of increasing size such that one can be confident theoretical errors are understood. A system with these properties could produce calculations in the physical limit for the first time from a real quantum computer.


A nearest-neighbor qubit connectivity is well-suited to simulating the $\mathbb{Z}_2$ gauge theory --- see, e.g. the Google Sycamore chip \cite{Arute2019} for a hardware realization.
While many $(1+1)d$ theories have been simulated on quantum devices (e.g. ~\cite{Martinez:2016yna,Kokail:2018eiw,Klco:2018kyo,Lamm:2018siq,Macridin:2018gdw,Klco:2019evd,Gustafson:2019mpk,Gustafson:2019vsd,Kreshchuk:2020aiq}), simulations of $(2+1)d$ QFT have never been --- in general the coherence requirements push these calculations into the era of quantum error correction (QEC) \cite{Shor1995}.
For near-term devices, it is important to simulate the impacts of quantum noise in order to understand how hardware improvements may enable beyond-classical computations of scientific interest even before the advent of full QEC.

The classical computing resources required to support research in quantum computing grow dramatically with problem size.
Clever techniques may sometimes keep scaling sub-exponential but in general this is impossible to avoid.
This necessitates flexible, large-scale computing resources and specialized hardware to solve problems of real scientific interest.

While large-scale quantum computing hardware is still years from general availability, quantum computing simulation with qsim \cite{quantum_ai_team_and_collaborators_2020_4023103} and Cirq \cite{cirq_developers_2021} are available for researchers exploring quantum programs on Google Cloud Platform (GCP) \cite{cirqqsim}.

Here we study the simulation of a $\mathbb{Z}_2$ gauge theory by simulating quantum circuits at a variety of increasing lattice sizes and noise levels to test an extrapolation to the physical limit. 
This is a challenging computational problem both in the scale of the number of simulations required and in the high-memory and size of the calculations, particularly for large lattices.
A six by six lattice is essentially at the bleeding edge of what is possible for classical computers for this problem.

This investigation will be the first practical demonstration of whether proper quantum dynamical simulations of a $\mathbb{Z}_2$
can be carried out in the near future in experimental quantum processors.
Additionally, this work establishes a lower boundary for quantum advantage in QFT simulations.
We also provide a template for demonstrating the feasibility of large-scale quantum simulation problems.
We expect demonstrations of this type will be an important ``gating step'' when running applications on real quantum computing platforms with the potential to address beyond classical problems.
Time on beyond-classical quantum resources is too scarce and valuable to deploy on problems that have not demonstrated both the requisite hardness and the ability to run on quantum resources.

\section{Theory}
The lattice field theory (LFT) program initiated by Wilson~\cite{Wilson:1974sk} has been successful in the study of nonperturbative quantum field theory. In order to render a QFT finite, LFT places the theory on a lattice in a finite volume.  Fields are placed either on the sites within the box, or on the links between them.  In this way, the infinite degrees of freedom of the QFT are rendered finite and can be simulated on a computer by sampling from the exponential of the action $S$.  In order to recover the true QFT results, calculations at different lattice spacings and volumes are performed and then extrapolated to the physical limit where both the finite volume and lattice spacing cutoff regulators have been removed. Unfortunately, for problems involving dynamics~\cite{Alexandru:2016gsd, Hoshina:2020gdy} or finite-density ~\cite{Gibbs:1986xg,Gibbs:1986ut,Troyer:2004ge,Alexandru:2018ddf} this method requires  exponential classical computational resources due to \emph{sign problems}. Such exponential costs can be avoided by using quantum devices.

Instead of using the action $S$, a more natural formulation of QFT for quantum devices is the Hamiltonian $\hat{H}$, which can then be used to time evolve a state using $\hat{U}(t)=e^{i\hat{H}t}$. For lattice gauge theories, the most commonly used $H$ is the Kogut-Susskind Hamiltonian~\cite{PhysRevD.11.395}.  In the case of $\mathbb{Z}_2$ gauge theory, it is
\begin{equation}
    \label{eq:GaugeHamiltonian}
    \hat{H}_{\text{gauge}} = -\gamma\left[\frac{1}{\beta_H} \sum_{i \in \text{links}} \hat{\sigma}^x_i- \beta_H \sum_{s} \prod_{i \in s} (\hat{\sigma}^z_i)^{\otimes}\right],
\end{equation}
where the two terms are analogous to the electric and magnetic terms found in electrodynamics -- U(1) gauge theory.
While this expression generalizes to other gauge theories, it requires a four-qubit operation.  This can be avoided by using the dual representation -- the transverse field Ising model~\cite{Wegner:1984qt,Kogut:1979wt,Yamamoto:2020eqi, Mathur_2016} -- which also reduces the number of qubits by half.
\begin{equation}
    \label{eq:DualHamiltonian}
    \hat{H}_{\text{dual}} = - J \sum_{\vec{n},\hat{\mu}} \hat{\sigma}^x_{\vec{n}}\hat{\sigma}^x_{\vec{n}+\hat{\mu}}- \Gamma \sum_{\vec{n}} \hat{\sigma}^z_{\vec{n}}\equiv H_K+H_V,
\end{equation}
where the sum over $\vec{n}$ ranges over the $N_s^2$ plaquettes, $\hat{\mu}=[\hat{x},\hat{y}]$ indicates spatial direction, $J = \frac{1}{\beta_H}$, $\Gamma = \beta_H$, and $\beta_H$ is related to the physical lattice spacing $a_s$. 

Current and near-term quantum devices have few qubits and large noise which in principle will limit calculations to small volumes and large $a_s$.  Whether simulations are sufficient for the physical limit depends on the size of the theoretical errors. Formally, these results should be equal to those obtained from an $S$ in the limit where the temporal lattice spacing is taken to zero. We can use this to leverage classical LFT Monte Carlo simulations to estimate the theoretical systematic errors from using current quantum devices~\cite{Carena:2021ltu} by computing results much closer to the physical limit. Obviously, in order to make these estimates the observable investigated must not have a sign problem, otherwise the classical simulations will be poor.  For this reason, the mass of the lowest-energy glueball state is chosen. Better candidates for near-term quantum advantage~\cite{Cohen:2021imf} are ones where the classical simulations are obstructed by sign problems, but the quantum simulations should have a similar difficulty and theoretical errors to the lattice glueball mass.

\begin{figure}[ht]
    \centering
    {\includegraphics[width=\linewidth]{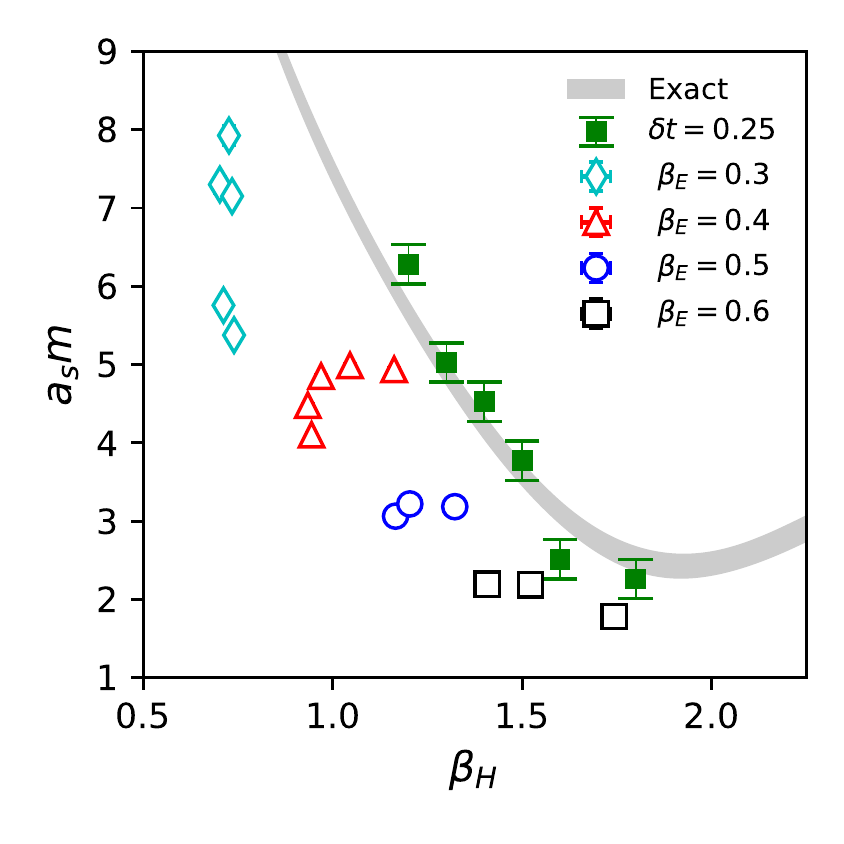}}
    \caption{Comparison of the lattice glueball mass $a_s m$ as a function of $\beta_H$ obtained from: (grey band) extrapolating the exact diagonalization of $\hat{H}$ from smaller volumes, (open symbols) classical simulations at fixed $\beta_E$ and varied $\xi$, (closed symbols) quantum simulations at fixed $\delta t$ for various $\beta_H$.}
    \label{fig:trotter66}
\end{figure}

For these calculations, the anisotropic Wilson action~\cite{Wilson:1974sk} was used:
\begin{equation}
    \mathcal{S} = -\beta_E / \xi_0 \sum_{{i}\in s} U_{{ss}} - \beta_E \xi_0 \sum_{i\in t} U_{{st}} 
\end{equation}
using plaquettes $U_{ij}$ which are oriented in the space-space or space-time planes with a coupling $\beta_E$ and a bare anisotropy factor $\xi_0$.  Lattice renormalization effects change the true anistropy of the lattice from $\xi_0$ to $\xi$.  As $\xi\rightarrow\infty$, for fixed $a_s$, the classical action results should converge to those obtained from the quantum simulations of the Hamiltonian. To check this, we can relate finite $\xi$ classical results to quantum simulation via the relation:
\begin{equation}
    \beta_E / \xi = \sqrt{\beta_H} e^{-\beta_E \xi}
\end{equation}
This convergence between the classical and quantum simulations is demonstrated in Fig.~\ref{fig:trotter66} for the case of our largest $6\times6$ square spatial lattice where the classical results (denoted by their value of $\beta_E$) approach the exact Hamiltonian result as $\xi$ is increased.
In addition, Trotterized real-time quantum results for $\delta t = 0.25$ are shown as well. Similar to the classical results. The Trotterized quantum results will only agree with the grey band in the limit that $\delta t$ approaches zero.

\begin{figure*}
\includegraphics[width=\textwidth]{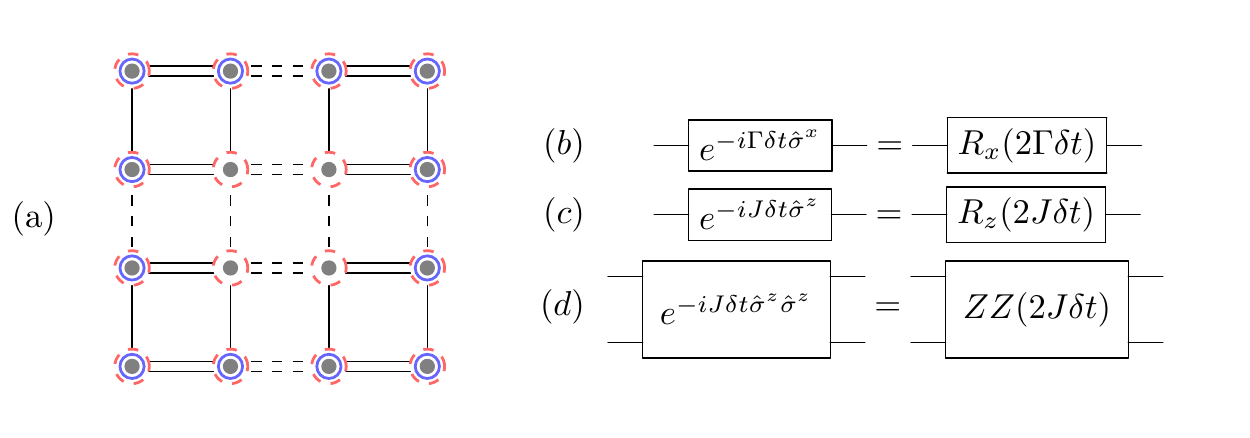}
\caption{Diagramatic depiction of the Trotter Circuit (a), and the gates in the Trotter circuit (b), (c), and (d). The different markers indicate the different gates and order that are applied in the Trotter circuit. Dashed red circle: $R_x(2 \Gamma \delta t)$ rotation layer 1; Solid blue circle: $R_z(2 J \delta t)$ rotation layer 2; Lines connecting qubits correspond to the rotation $e^{-i J \delta t \hat{\sigma}^z\hat{\sigma}^z}$ applied on layer 3 (single solid line), layer 4 (single dashed line), layer 5 (double solid line), and layer 6 (double dashed line). The ancillary qubit used for computing a non-Hermitian observable is not shown. }
\label{fig:TrotterDiagram}
\end{figure*}
By studying the difference between the lattice results and the continuum on the classical side, estimates of the theoretical errors from an analogous quantum simulation can be found.
These estimates direct the choice of $J,\Gamma$ and Trotter step for the quantum simulations and determining when continuum extrapolation may be viable.

For the quantum simulation, we use the observable~\cite{Gustafson:2020yfe}
\begin{equation}\label{eq:observable}
    \mathcal{C}_{i,s}(t) = \langle \Omega | \hat{U}^{\dagger}(t) \hat{X}_i \hat{U}(t) \hat{X}_{s} |\Omega\rangle.
\end{equation}
The subscript $s$ indicates the location on the lattice that we want the source operator $\hat{X}$ to act on, while $i$ iterates over a subset of sites in the lattice. $|\Omega\rangle$ is an approximate ``ground state" which we choose to be $|0...0\rangle$.
This correlator can be expanded in the eigenbasis of the time evolution operator,
\begin{equation}
    \hat{U}(t) = \sum_{E} |E\rangle \langle E| e^{-i t E}.
\end{equation}
Using this expansion, we can write the operator $\mathcal{C}_{i,s}(t)$ as
\begin{equation}\label{eq:fourier}
    \mathcal{C}_{i,s}(t) = \sum_{\lbrace E_k \rbrace, \lbrace E_m \rbrace} A_{i,s}(E_k, E_m) e^{i t (E_k - E_m)},
\end{equation}
where the factors $A_{i, s}(E_k, E_m)$ account for the various inner products from the eigenbasis expansion. Since the time dependence of the correlator is governed by the differences in energy levels, we can extract particle state energies given by energy gaps above the ground state. These energy levels can be extracted by performing a fast Fourier transformation on the time series data to generate a frequency spectrum. This frequency spectrum will have peaks at energy differences $|E_k - E_m|$ that most significantly drive the dynamics of the simulations. Provided the overlap of the state $|\Omega\rangle$ with the ground state is sufficient enough, the energy differences should be most strongly governed by excitations above the ground state $E_k - E_0$. With the set of operators that we chose, the strongest excitation (largest peak in Fourier spectrum) will correspond to the glueball mass (mass gap), $E_1 - E_0$.

Time evolving a quantum system requires the unitary operator of $\hat{U}(t)=e^{-i\hat{H}t}$ which cannot in general be efficiently constructed on a quantum computer. Instead, it must be approximated. A common method is Trotterization, whereby the time $t$ is subdivided into $N$ smaller steps of size $\delta t$ and $\hat{U}(t)\approx (e^{-i\hat H' \delta t})^N$ with an approximate Hamiltonian $\hat H'$. 
The eigenvalues and eigenvectors of $\hat{H}'$ will only formally agree with the eigenvalues and eigenvectors of Equation \ref{eq:DualHamiltonian} in the limit that $\delta t$ goes to zero.
The quantum circuits for the Trotterized time evolution of the $\mathbb{Z}_2$ gauge theory are shown in Fig. \ref{fig:TrotterDiagram}

\section{Noise model}

\subsection{Physics details}

Since fault-tolerant quantum computation using dozens of qubits is currently unavailable, we implement several simple noise models to study the parameter regime in which this problem might be feasible on NISQ \cite{Preskill_2018} devices. Crosstalk is expected to be a significant source of error in superconducting qubit processors, and we simulate this effect by implementing the unitary noise model for crosstalk in $\sis$ gates of ref. \cite{McKay_2019}. In addition we simulate the combined effects of gate infidelity and qubit error (e.g. $T_1$ and $T_2$ decay) using local depolarizing channels.  

\subsubsection{Local depolarization}

We study the effects of incoherent noise by simulating local symmetric depolarizing noise. A depolarizing channel acts on an $n$-qubit subsystem ($2^n$-dimensional system) via the completely positive, trace-preserving map

\begin{equation}\label{eq:depol_1}
    \mathcal{D}_n[\epsilon](\rho) = (1 - \epsilon) \rho + \frac{\epsilon}{2^n} I
\end{equation}

We define a probability mass function over $\{0, 1, 2, 3\}^n$ given as
\begin{align}
p(j) = \begin{cases}
    1 - (1-4^{-n})\epsilon & j = 0^n \\
    4^{-n} \epsilon &  \text{else}
\end{cases}
\end{align}

An operator sum representation for Equation~\ref{eq:depol_1} is then given as a sum over $n$-local Pauli operators $P_j = \sigma_{j_1} \otimes \dots \otimes \sigma_{j_n}$ as:
\begin{equation}\label{eq:depol_kraus}
    \mathcal{D}_n[\epsilon](\rho) = \sum_{j \in \{0,1, 2, 3\}^n} p(j) P_j \rho P_j
\end{equation}

Our noise model incorporates this type of error by applying $\mathcal{D}_1[\epsilon_1]$ to each qubit following each single-qubit gate and $\mathcal{D}_2[\epsilon_2]$ to each two-qubit subsystem following each two-qubit gate, taking $\epsilon_2 = 10 \epsilon_1$ as which is a reasonable ratio for superconducting qubit devices.
For smaller simulations we use an ancillary qubit to compute $\mathcal{C}_{i,s}(t)$ in a single simulation, however we do not apply noise to the ancillary qubit since our circuits would be executed on near-term devices without the ancilla using the methods of ref. \cite{Mitarai_2019}. While it is rare to observe symmetric depolarizing noise in real devices, this noise model allows for probing very general noise effects in a relatively small parameter space. Furthermore, our noise model can provide insight for the use of techniques such as randomized compiling \cite{Wallman_2016} which combine randomized experiments such that the observed noise resembles local depolarizing noise. This technique has been demonstrated to be effective at modifying the noise behavior for the kind of circuit-based Hamiltonian simulation that we perform here \cite{PhysRevX.8.031027}.

\subsubsection{Two qubit gate crosstalk}

Crosstalk is expected to be a significant source of error in the implementation of two-qubit gates that will limit the scale of algorithms implemented on near-term hardware (e.g. \cite{Sarovar_2020}). We study the effects of crosstalk in the context of $\sqrt{\text{iSWAP}}$ gates according to the unitary map $\Lambda_{ZZ}( \rho) = U_{ZZ} \rho U_{ZZ}^\dagger$ on each 2-qubit subsystem, where the unitary error term is given by \cite{McKay_2019,Mundada_2019}
\begin{equation}\label{eq:zeta_noise}
    U_{ZZ}[\zeta] = \exp \left(-i2\pi \zeta T |11\rangle\langle 11| \right)
\end{equation}

This model has a fixed parameter $T$ corresponding to the duration of the $\sis$ gate (roughly $10^{-8}\, s$ on Google hardware) and a free parameter $\zeta$ that describes the crosstalk interaction strength arising from fabrication conditions and device parameters. To implement this model in simulation, we decompose each two-qubit gate in our circuit into two $\sis$ gates, each prepended by $U_{ZZ}[\zeta]$. This noise model for crosstalk is unitary and therefore does not require trajectory simulation in isolation, but will generally require trajectory simulations when combined with the depolarizing noise model.

\begin{table}[ht]
\centering
\begin{tabular}{c|c|c|c|c}
& \multicolumn{2}{c|}{noiseless} & \multicolumn{2}{c}{noisy} \\
\hline
Grid & 1q gates & 2q gates & 1q gates & 2q gates \\
3x3 & 18 & 12 & 342 & 96 \\
4x4 & 29 & 24 & 677 & 192 \\
5x5 & 42 & 40 & 1122 & 320 \\
6x6 & 57 & 60 & - & - \\
\hline
\end{tabular} 
\vspace{0.3cm}
\caption{Gate counts per Trotter step for each $\mathbb{Z}_2$ simulation. Noiseless figures assume hardware-native $\exp(i\theta Z_i Z_j)$ entangling gate. Noisy figures involve decomposition of each two qubit gate into $4$ $\sis$ gates and local rotations, insertion of unitary noise after each $\sis$ gate, and stochastic insertion of depolarization Kraus operations assuming $\epsilon_1= 5 \times 10^{-4},\,\epsilon_2 = 5 \times 10^{-3}$. Figures do not account for virtual $R_z$ gates typically implemented on currently available superconducting qubit hardware \cite{cirq_developers_2021,qiskit}. We do not perform noisy simulation on the $6 \times 6$ system }
\label{tab:circuit_sizes}
\end{table}

Table~\ref{tab:circuit_sizes} presents trotter step specifications for noiseless and noisy simulation of $\mathbb{Z}_2$ systems. For noisy simulations we decompose $\exp(i\theta Z_i Z_j)$ entangling gates into the gateset $\{ \sis, R_x, R_y, R_z\}$. Each decomposition results in $4$ $\sis$ gates and 24 local operations. Further optimization could reduce the gate count of this decomposition by approximately one half, and so the results of our analysis of noisy simulation are conservative. We perform this decomposition before inserting the noisy operations $\mathcal{D}_1[\epsilon_1]$ after each single qubit gate and $\mathcal{D}_2[\epsilon_2]$ and $U_{ZZ}[\xi]$ after each two-qubit gate. 

\subsection{Implementing the noise model using trajectory simulation}

The trajectory simulation feature for noisy circuit simulation in qsim is based on the Kraus operator sum representation of quantum channels. The effect of $\mathcal{D}_n[\epsilon]$ in Equation~\ref{eq:depol_kraus} is therefore approximated by first executing a series of independent state vector simulations, each of which stochastically applies an operator $P_j$ sampled according to $p(j)$ in place of applying $\mathcal{D}_n[\epsilon]$, and then combining the resulting state vectors as an evenly-weighted ensemble to produce a finite-trajectory approximation to $\mathcal{D}_n[\epsilon] (\rho)$. 

\section{Multi-node simulation}

\subsection{Distributing trajectories over multiple compute nodes}
To roughly approximate the effect of $\mathcal{D}_n[\epsilon]$ using trajectory simulation, 1000 iterations of the independent state vector simulations described in III.B were executed. This achieves an approximation with a statistical error of $\sim 1 / \sqrt{{\rm iterations}}$. The state vector simulations were distributed over multiple compute nodes for performance.

\begin{figure}
    \centering
    {\includegraphics[width=\linewidth]{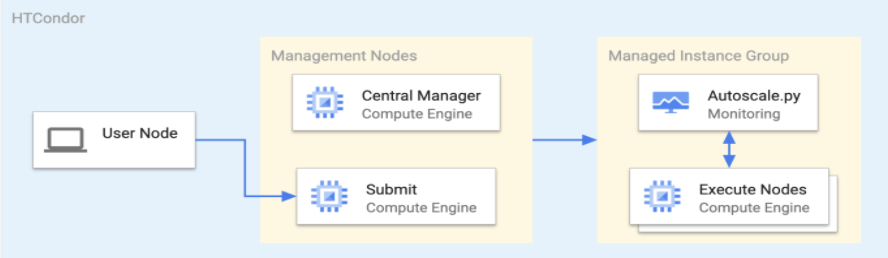}}
    \caption{HT Condor workflow for multinode simulation}
    \label{fig:HTcondor}
\end{figure}

An autoscaling workflow which uses HTCondor as a scheduler was used to distribute the simulations. The workflow creates the virtual machines required for the simulation and adds them to the managed instance group (MIG).  An MIG is a collection of virtual machine instances that are created from a common instance template. The autoscaling workflow adds or deletes instances from the managed instance group based on the group's autoscaling policy (see figure \ref{fig:HTcondor}). The autoscaling policy defined for quantum simulations makes a decision to increase or decrease the number of nodes in the MIG based on the number of pending jobs returned by the condorq command. The autoscaler is built to cope well with heterogeneous clusters (containing multiple machine types in multiple regions): it uses HTCondor matchmaking algorithms to require or prefer co-locating compute and storage for optimal performance.

\subsection{Speeding up noisy simulations by using GPUs}
GPU support was recently brought to qsim, and we test it at scale in this work. qsim performance depends on fast routines for gate fusion and matrix-vector multiplication. Matrix-vector multiplications are performed for gate matrices of size $ {2}^k \times {2}^k$ and state sub-vectors of size $ 2^k$, where $k$ is is the gate size in qubits. Such multiplications are performed in parallel for multiple sub-vectors. In the case of CPUs, the number of sub-vectors that can be processed in parallel is equal to the SIMD register size in floats. This is parallelized further using OpenMP. In the case of GPUs, the number of sub-vectors that can be processed in parallel is equal to the number of threads the streaming multiprocessors can run simultaneously.

The GPU implementation in qsim uses CUDA. The implementation efficiently utilizes the GPU compute  resources  and  memory  bandwidth.  Within a certain parameter space, qsim can be three to ten times faster on a GPU than a CPU. As a rule of thumb, there is a significant performance gain for noisy 23-30 qubit simulations.

Table \ref{tab:benchmarks_small} presents wall times for simulating $\mathbb{Z}_2$ systems of up to 26 qubits using different Google Cloud computing platforms. We ran noisy circuit simulations for $3\times 3$ and $4 \times 4$ grids using Nvidia-T4 GPUs, and $5 \times 5$ grids using Nvidia-V100 GPUs. Benchmark timings are best-of-3 computed using $\zeta=10^5$ and $\epsilon_2=0.005$ (defined in Equations \ref{eq:zeta_noise} and \ref{eq:depol_kraus} respectively), averaged over computing 20 Trotter steps for 50 trajectories and do not include overhead for data transfer into the GPU, which can dominate simulation time for small trajectory counts. The figures slightly underestimate the typical timing since initial Trotter steps are applied to a sparse state. Simulations on CPUs are performed using the $\textsc{qsim}$ denormals-are-zero option ('z' option) set to true. This option does not affect the GPU simulations.

\begin{table}[h]
\centering
\begin{tabular}{c | l | l | l }
Platform & $3 \times 3\,(10)$ & $4 \times 4\,(17)$ & $5 \times 5\,(26)$ \\
\hline
n1-standard-32 & 0.342 & 1.391 & 25.97 \\
c2-standard-30 & 0.252 & 0.935 & 24.84 \\
Nvidia-T4 &  0.344 & 0.962 & 11.15 \\
Nvidia-P100 &  0.310 & 0.962 & 10.33 \\
Nvidia-V100 &  0.367 & 1.017 & 3.872 \\
Nvidia-A100 &  0.384 & 1.090 & 3.220 \\
\hline
\end{tabular}
\vspace{0.3cm}
\caption{Wall time in seconds to simulate each grid (number of qubits), per 100 trotter step trajectories for a single choice of $\beta_H$ (defined in Equation~ \ref{eq:GaugeHamiltonian}) and $\delta t$ (defining the trotterization timestep).}
\label{tab:benchmarks_small}
\end{table}
\section{TPUs}

Tensor processing units (TPUs) \cite{jouppi2017indatacenter} are specialized Machine Learning ASICs that greatly accelerate training and inference of large ML models \footnote{Researchers may find more information about access to TPUs at \url{https://sites.research.google/trc}}.
Recent work has shown that they can also be repurposed for other tasks, such as the simulation of quantum circuits described in this work and large-scale quantum chemistry computations \cite{tpu_qchem}.
TPU simulations for 36 qubits were run on a 512-core configuration of TPUv3, equipped with 8TB of high bandwidth memory. The simulation code was written in JAX \cite{jax2018github}.
%
A single 100-Trotter step run of the circuit (including measurements) took on the order of 630 seconds, with approximately 1/3 of the time dedicated to computing observables. Spatial symmetry was exploited to reduce computation times of observables by roughly 1/2.
We estimate a factor of roughly 105 total speed up for the full quantum simulation workload (physics parameter scan and computation of all observables) on the TPUv3 platform over an OpenMP parallelized qsimcirq simulation on an m1-ultramem-160 shared-memory machine with 160 virtual CPU cores. 

\begin{table}[h]
\centering
\begin{tabular}{c  | l }
Platform  & $6\times 6$ \\
\hline
m1-ultramem-160 (qsimcirq) & 470 hours \\
m1-ultramem-160 (qsim) & 295 hours \\
TPUv3-512  & 4.5 hours \\ \hline\end{tabular}
\vspace{0.3cm}
\caption{Comparison of total runtime on TPUs to runtime estimate for m1-ultramem-160 to compute the observables after each of 100 trotter steps for a set of 25 choices of $(\beta_H, \delta t)$. The qsimcirq implementation simulates a 37 qubit system to compute observables for the $6 \times 6$ grid using an ancilla. The qsim implementation estimates the time necessary to simulate two 36-qubit state vectors, $|\psi_1\rangle = U(t)|\Omega\rangle$ and $|\psi_2\rangle = U(t) X_s |\Omega\rangle$, and then compute $\mathcal{C}_{i,s}(t) = \langle \psi_1| X_i |\psi_2\rangle$ of Equation~\ref{eq:observable} for each $i=1\dots 21$ (yielding the remainder of observables by spatial symmetry). The TPU simulation implements this strategy as well, but this technique is currently unavailable using qsimcirq.
We omit comparison to GPU simulation which is currently not supported for 37 qubits, due to memory limitations of existing GPU architectures.}
\label{tab:benchmarks_large}
\end{table}

Table \ref{tab:benchmarks_large} presents wall times comparison for simulation of a $6 \times 6$  $\mathbb{Z}_2$ system using TPU resources (a TPU v3-512 pod-slice with 512 TPU cores) versus an ultra high memory CPU cloud resource (160 virtual CPU cores, 3.8 TB memory).




\section{Data analysis}

\subsection{Noisy simulation study}

To study the effects of simulated noise on the $\mathbb{Z}_2$ simulation we performed a sweep over physical parameter space and noise parameter space. Each choice of tuple $(\beta_H, \delta t, \xi, \epsilon_2)$ characterizes a single choice of physical and noise parameters that we simulated for 50 trotter steps using 1000 trajectories, requiring $1.25 \times 10^6$ total trotter step trajectories per point in parameter space. We used the physical parameters $\beta_H \in [1.4, 1.6, 1.8]$ and $\delta t=0.25$ in simulations for 50 trotter steps. The values of $\beta_H$ were chosen so that finite volume effects corresponding to $a_s(\beta_H) N_s$  were balanced against wanting simulations close to the continuum limit $\beta_H \sim 1.85 - 2.0$. The choice of $\delta t = 0.25$ was done so that energy was small enough to extract a meaningful enough signal and $n_t$ would provide a reasonable resolution on the energy, $\approx 0.1 / a_0$. 

For each choice of physical parameters $(\beta_H, \delta t)$ we performed simulations on each $n \times n$ grid sweeping over the following noise parameters

\begin{align}
    \epsilon_2 &\in \begin{cases} [0, 1, 2, 3, 4, 5] \times 10^{-3} & n=3 \\
    [0, 0.5, 1, 1.5, 2, 2.5, 3.0] \times 10^{-3} & n=4,5 
    \end{cases}
    \\
    \zeta &\in [1.5, 3.0, 4.5, 6.0, 7.5] \times 10^5
\end{align}

\begin{figure*}[ht]
    \centering
    \includegraphics[width=\textwidth]{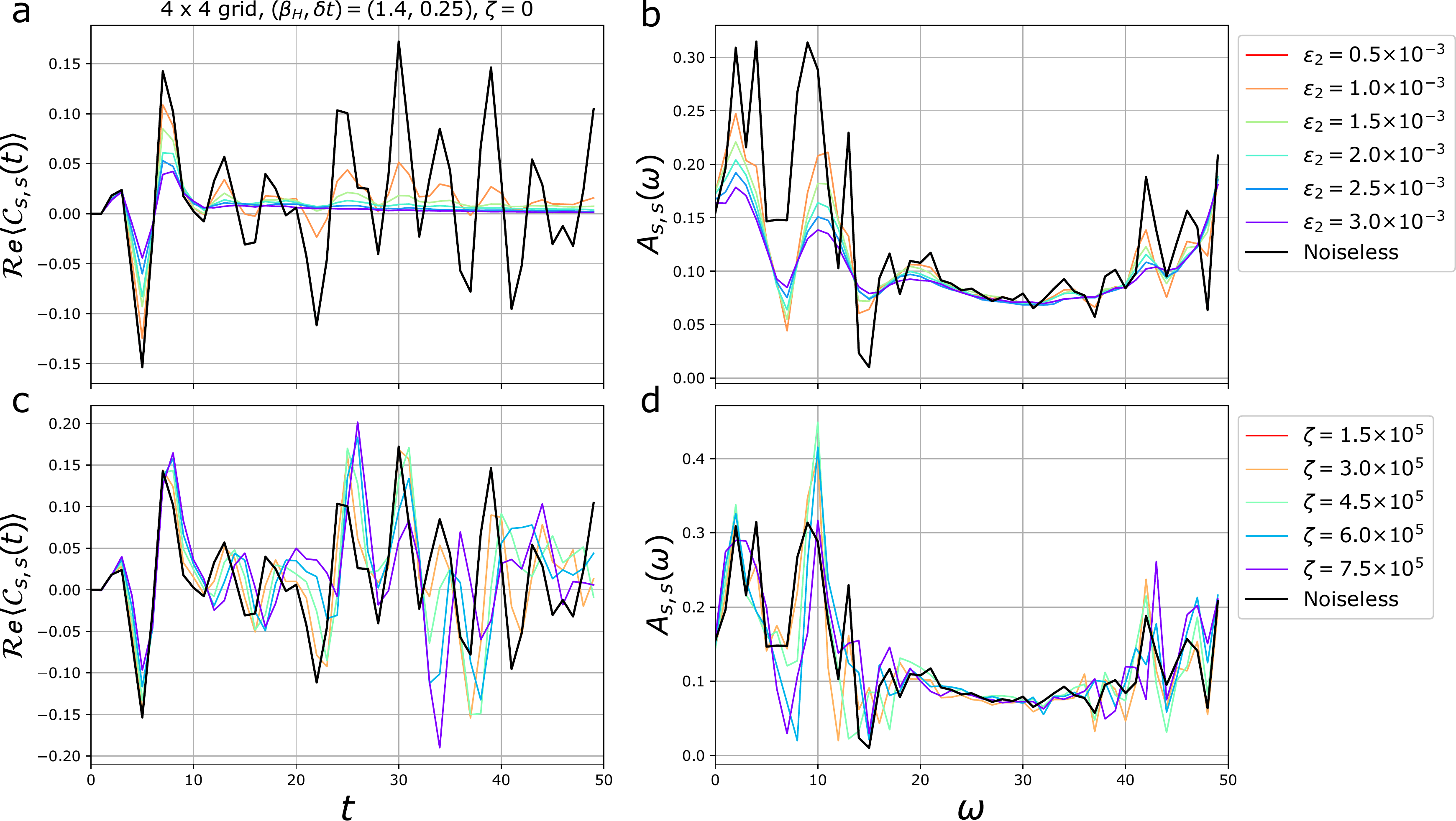}
    \caption{(a) Time series and (b) Fourier transform for increasing $\epsilon_2$ in the depolarizing model with fixed $\zeta=0$ on a $4 \times 4$ grid. The effect of depolarizing noise is to dampen the time series signal, which both flattens the existing spectrum and introduces higher frequency modes. The maximum $\epsilon_2=$ corresponds to an error of $236\%$ in the computed glueball mass and tends to increase with $\epsilon_2$. (c) Time series and (d) Fourier transform for increasing $\zeta$ with fixed $\epsilon_2=0$. The crosstalk error model tends to distort the spectrum in an unpredictable manner. The maximum $\zeta=7.5\times 10^5$ corresponds to an error of $37\%$ in the computed glueball mass, highlighting that the computed mass is highly non-monotonic with respect to the crosstalk noise.}
    \label{fig:eps_zeta_sweep_sample}
\end{figure*}

Figures~\ref{fig:eps_zeta_sweep_sample}a-b demonstrates the time dependence of the quantity $\mathcal{C}_{s, s}(t)$ (Equation~\ref{eq:observable}) and its Fourier coefficients $A_{s,s}(\omega)$ (Equation~\ref{eq:fourier}) for $(\beta_H, \delta t, \zeta)=(1.4, 0.25, 0)$ with increasing $\epsilon_2$. The time series is strongly damped in $\epsilon_2$ due to the tendency of depolarizing noise to contract the reduced density matrix over each qubit to the maximally mixed state $I/2^n$. This has the effect of suppressing in the spectrum $A_{s,s}$, which distorts the glueball mass computed from the  gap between the two lowest frequencies in Equation~\ref{eq:fourier}.

Figures~\ref{fig:eps_zeta_sweep_sample}c-d demonstrate the effects of increasing $\zeta$ for $(\beta_H, \delta t, \epsilon_2)=(1.4, 0.25, 0)$ on the $\mathbb{Z}_2$ simulation. Unlike the case of depolarizing noise, there is no clear trend in the behavior of the quantities $\mathcal{C}_{s, s}(t)$ and  $A_{s,s}(\omega)$ with increasing $\zeta$, highlighting the need to explore the effects of coherent noise via circuit simulations. We further note that simulated crosstalk noise has the largest effect on smaller $\beta_H$ values. This is possibly due to the sensitivity of small $\beta_H$ simulations to higher frequency modes, which will be more easily corrupted by the presence of coherent noise involving fixed rotation angles.

\begin{figure}
    \centering
    \includegraphics[width=\columnwidth]{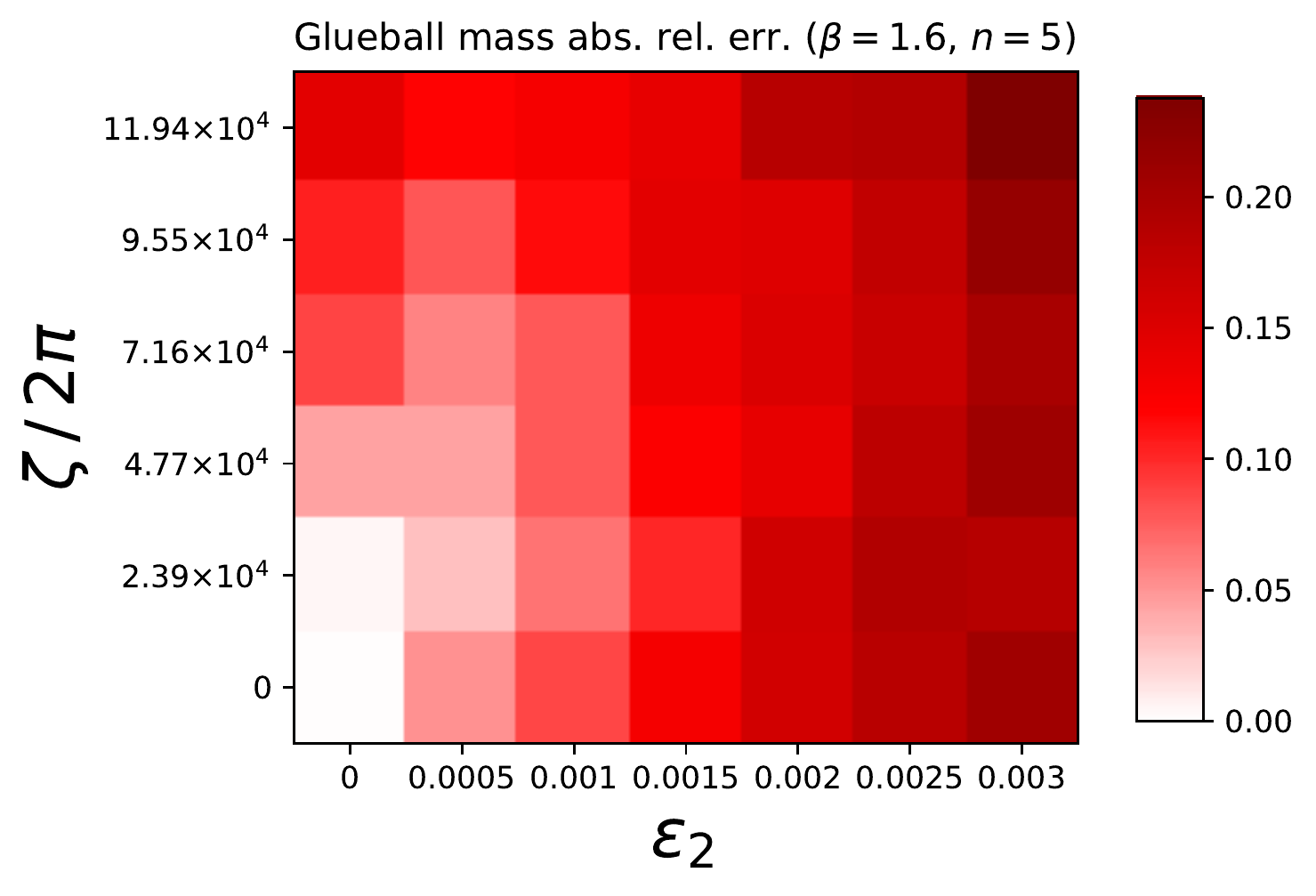}
    \caption{Relative error in computed glueball mass as a function of $(\epsilon_2, \zeta/2\pi)$ noise strength shows a narrow regime in which the mass can be computed with high accuracy for the $5 \times 5$ grid. For comparison, ref. \cite{Mundada_2019} describes how $\zeta/2\pi=2.26 \, \times 10^6 \,s^{-1}$ results in crosstalk becoming the dominant source of error for transmon qubits with $T_1=15.2 \mu s$, $T_2=4.2 \mu s$ coupled via a bus cavity (which differs from the architecture employed in the Google Sycamore chip). The contents of this plot required roughly 500 GPU-hours of simulation time using Nvidia V100 GPUs.}
    \label{fig:n3_err_map}
\end{figure}

Figures~\ref{fig:eps_zeta_sweep_sample}a-d demonstrate the effects of either noise model in isolation on the observed time series. To determine the effects of the combined noise model, we analyzed the results of each noisy simulation and compared the computed glueball mass to that of the corresponding noiseless simulation. Figure~\ref{fig:n3_err_map} shows an example outcome for a sweep over the noise parameter space for a $3 \times 3$ grid with $(\beta_H, \delta t) =(1.6, 0.25)$. The complete parameter space sweeps were simulated in 108 GPU-hours for the $3\times 3$ grid and 360 GPU-hours for the $4\times 4$ grid (both using Nvidia T4 GPUs), and 1500 GPU-hours for the $5\times 5$ grid distributed across 64 Nvidia V100 GPU  instances using the multinode qsim implementation.

\subsection{Error budget estimate}

For quantum simulations on a single device with fixed coupling $\beta_H$, the two dominant sources of systematic error are the theoretical finite volume errors and the quantum noise.  The finite volume errors can be mitigated by performing calculations at multiple volumes and extrapolating.  

\begin{figure}
    \centering
    \includegraphics[width=\columnwidth]{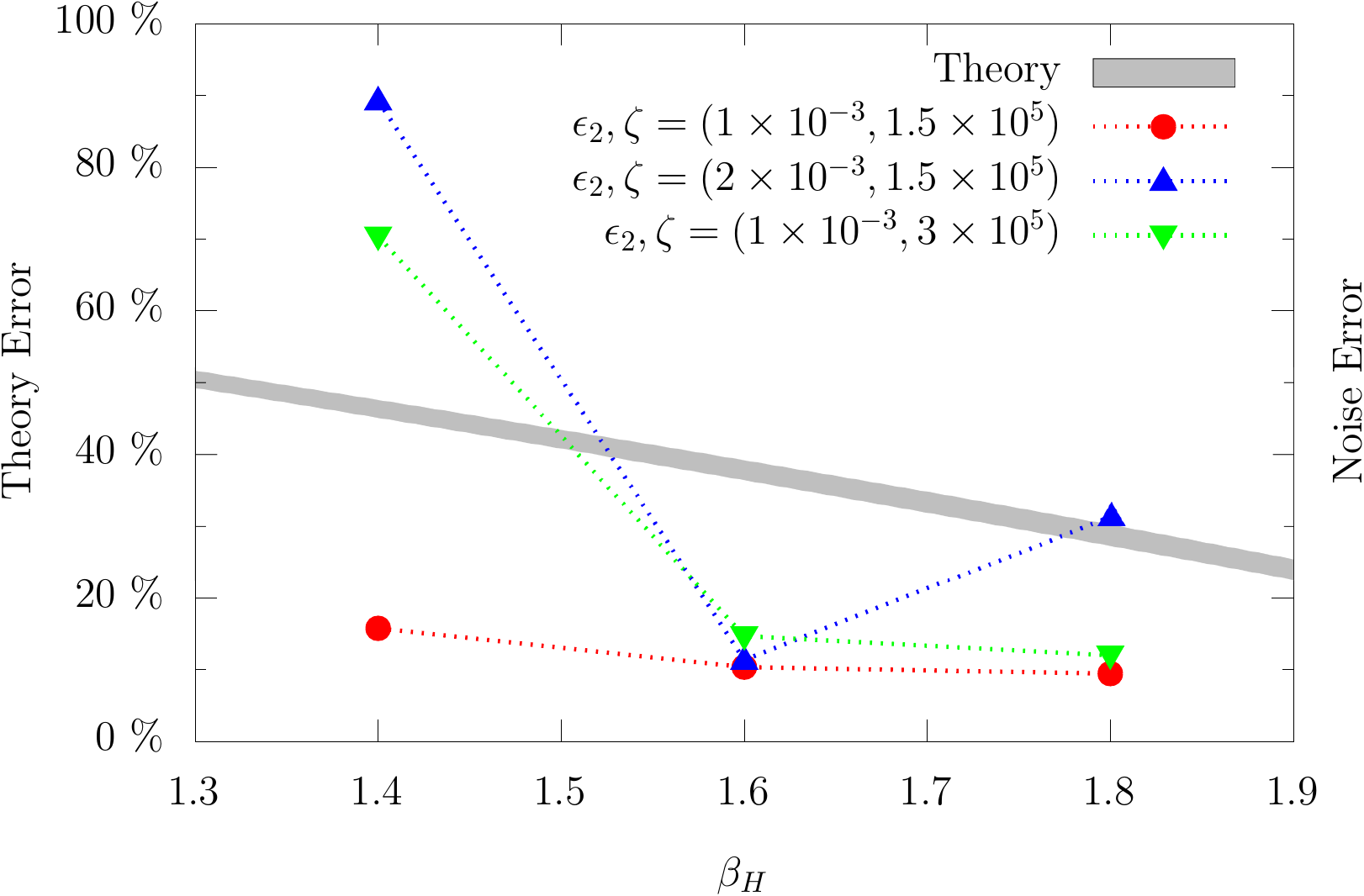}
    \caption{Sources of relative systematic error in computed glueball mass as a function of $\beta_H$. The gray band indicates the estimated theoretical errors from extrapolating with $3\times$, $4\times4, 5\times5,$ and $6\times6$ classical lattices. The error from noise are shown for different fiducial noise models.}
    \label{fig:err_comp}
\end{figure}

Using our classical results, we have estimated the discrepancy from the finite volume by extrapolating $3\times3$, $4\times4$, $5\times5$, and $6\times6$ results together, and comparing to calculations performed on much larger lattices where finite volume effects are negligible~\cite{Agostini:1996xy}. These errors are plotted in Fig.~\ref{fig:err_comp} where they are compared to the relative error from quantum noise.  Efficient use of quantum resources occurs when the theoretical and noise errors are comparable. The overall relative error that could be tolerated for quantum advantage depends strongly on the prior knowledge of a given observable. For the glueball mass, as previously discussed, the total error would need to be sub-percent level because it can be precisely computed classically.  While this error budget is only for this observable, it should be qualitatively similar for others such as transport coefficients, where the total error could be as large as 100\% and still be competitive with current classical results.

\section{Conclusion}

We demonstrated the usefulness, power, and flexibility of GCP for studying quantum simulation of quantum field theories.
While some GCP hardware used here such as the TPUs are not yet broadly commercially available, we demonstrated their potential for addressing questions of scientific interest in quantum information science in the near term.
Our simulations and analysis provide important context --- both at the theoretical level and in terms of quantum noise tolerance --- for using increasingly available quantum hardware to study quantum field theories and other fundamental physics problems.

The context of feasibility for physics on NISQ era hardware is dependent on the accuracy of the observable of interest.
The glueball mass may be obtained accurately from classical Mont\'e Carlo (at the sub-percent level).
Our simulation of quantum circuits, using lattice sizes up to $6 \times 6$, cannot provide an uncertainty comparable to this.
For observables for which the classical Mont\'e Carlo fails we expect similar theory errors.
Therefore, unless high 
error is acceptable, the requisite exact circuit calculation is beyond our ability to simulate classically, and thus requires appropriate quantum computing hardware.
However, it remains possible that some subset of the observables, such as transport coefficients, could be competitive --- further research is necessary.

\section{Reproducibility}

A copy of all of the circuits and GCP job submission code is available at \url{https://github.com/Fermilab-Quantum-Science/Z2Sim-public}.
These circuits use Cirq \cite{cirq_developers_2021} and qsim \cite{quantum_ai_team_and_collaborators_2020_4023103}, which are open-source software.
The specific code framework used to generate code for the TPUs is not public at this time, but the same results may be derived on a sufficiently powerful computing platform using the circuit code provided.

\section*{Acknowledgments}

We thank Google Cloud for supporting this research and offering time on the required machines. We thank Alan Ho for his important role in creating this project.
This work was partially supported by the DOE/HEP QuantISED program grant HEP Machine Learning and Optimization Go Quantum, identification number 0000240323. This work was partially supported by the DOE through the Fermilab QuantiSED program in the area of "Intersections of QIS and Theoretical Particle Physics." 
This manuscript has been authored by Fermi Research Alliance, LLC under Contract No. DE-AC02-07CH11359 with the U.S. Department of Energy, Office of Science, Office of High Energy Physics.



\bibliographystyle{IEEEtran}
\bibliography{biblio}
%
%
%

\end{document}